\documentclass[conference]{IEEEtran}
\usepackage{cite}

\ifCLASSINFOpdf
  \usepackage[pdftex]{graphicx}
   \graphicspath{{C:/Users/Administrator/Desktop/latex/eps}}
  \DeclareGraphicsExtensions{.pdf,.jpeg,.png,.eps}
\else
\fi
\usepackage{CJK}
\usepackage[numbers,sort&compress]{natbib}
\usepackage{cite}
\usepackage{amsmath,amssymb,enumerate}
\usepackage{cite,graphics,graphicx}
\usepackage{graphicx}
\usepackage{subfigure}
\usepackage{listings}
\usepackage{makeidx}
\usepackage{hyperref}
\usepackage{algorithm}
\usepackage{algorithmic}
\usepackage{float}
\usepackage{multirow}
\usepackage{amsmath}
\usepackage{amssymb}
\usepackage{xcolor}
\usepackage{epsfig}
\usepackage{graphics}
\usepackage{listings}
\usepackage{algorithm}
\usepackage{algorithmic}
\hypersetup{linkcolor=black, %
            citecolor=black, %
             pdftitle={Title}, %
            pdfauthor={Name}, %
           pdfsubject={Subject}, %
          pdfkeywords={Key words}}
\begin{document}

%
\title { A Tag Identification Approach Based On Fragile Watermark}

\author{\IEEEauthorblockN{Jianbiao Lin{$^{1}$},
Ke Ji{$^{2}$},
Hui Lin{$^{3}$},
Enyan Wu{$^{4}$},
Xin Gao{$^{4}$}
}

\IEEEauthorblockA{
1.2.Computer Science and Engineering Department\\ Sichuan University Jinjiang College, Penshan 620860, China\\
3.College of Civil Engineering\\ Sichuan University Jinjiang College, Penshan 620860, China\\
4.College of Economics\\ Sichuan University Jinjiang College, Penshan 620860, China\\
Email: zeroyuebai@hotmail.com}
}


%


\maketitle

\begin{abstract}
 This paper proposes a tag identify approach based on fragile Watermark that based on  Least significant bit of the replacement that we first use a special way to  initialize the cover to ensure that we can use  random positions to embed the information of tag. Using this  way enhance the security of other to get the right information of this tag. Finally as long as the covered information can be decoded, the completeness and accuracy of the tag information can be guaranteed. the result of simulation experiment show that this approach has high sensitivity and security .\\\\
 keywords- watermark£¬security£¬robutness£¬fragile

\end{abstract}


\section{Introduction}
With the rapid development of society and economy. These advances have also made it possible to copy and modify the tag of other company to gain benefit. There the tag protection for company has received widespread attentions. Fragile watermarking technique, as a kind of new technique of tag protection ,has been extensively researched, and many fragile watermarking schemes has been proposed. In general ,a fragile watermarking is sensitive to the information that is to say we can use fragile status to judge whether the information is changed. According to the different type of protected tag, fragile watermarking can be divided into image watermarking ,audio watermarking, video watermarking, etc. This paper focus on image watermarking, which has two conflicting requirements: imperceptibility and robustness. The conflicting point is how to improve the watermark¡¯s ability against the attacks while influencing as less as possible the original image. Using low Significant bits algorithm is a good way to embed the tag.

This paper present a approach that embed a tag in a picture. due to the sensitive of the information ,when it be attacked. This approach using a random string to decide where the tag will be insert. Further more this approach is a bind watermarking technique, which means that the extraction of the tag does not require the original. The following sections describe how to embed the tag to cover in detail, the simulation experiment shows that this fragile watermarking approach is safe, reliable and has a good invisibility and some good reflect about  what was attacked.

\section{ Attack schemes and evaluation benchmark}\label{SEC:  Attack schemes and evaluation benchmark}

Similar cryptography, digital watermarking technology in practical applications is bound to suffer a variety of attacks .People curious about new technologies, piracy will bring huge profits to become motivated attacks (malicious); And digital products in storage, distribution, printing, scanning process, what can introduce a variety of distortion (unintentional attacks). Therefore watermarking algorithm designer to deal with a series of tests watermark image, watermark robust performance evaluation, according to the assessment process and the emergence of various kind of problem accordingly modified algorithm. currently there have been many attacks specifically assess the watermark quality software, such as Mo saic, D igim arc, Su reSign, St irM ark, etc.Similar cryptography, digital watermarking technology in practical applications is bound to suffer a variety of attacks people curious about new technologies, piracy will bring huge profits to become motivated attacks (malicious); And digital products in storage, distribution, printing, scanning process, can introduce a variety of distortion (unintentional attacks). Therefore watermarking algorithm designer to deal with a series of tests watermark image, watermark robust performance evaluation, according to the assessment process and the emergence of various kind of problem accordingly modified algorithm. Currently there have been many attacks specifically assess the watermark quality software, such as Mo saic, Digim arc, SureSign, StirM ark, etc.For example StirM ark, the image it tiny, naked eye can not detect the geometric distortion, such as stretching, cropping, moving, bending, etc., and after the image has been imitated high quality printers, high-quality scanner, and the image is introduced error. Furthermore, for each pixel of the image are incorporated tiny, what is uniformly distributed error value meet, and to mimic a slight error in the scanner and the display device, which is need nonlinear analog / digital converters produced.

Fragile image watermarking is an important branch of image watermarking technology.In addition to the basic features have a watermark, but also with data integrity and validity of annotation capabilities, as well as data breaches and attacks positioning analysis capabilities, and for different applications have different robustness.

Image watermarking algorithm test object should be drawn from the perspective of different standard image signal processing point of view, and the picture should have the following characteristics: a textured / smooth region; obvious edge; can be acute treatment, obfuscation and brightness / contrast adjustment.

\subsection{Initial the image}\label{SSEC: Initial the image}

In order to ensure that we can using random way to decide where the tag will insert, we must Initial the image to let the least significant become one if the least is zero else stay the same. We also need to make the lower  Significant of the tag become zero, and let the higher bit of the tag become the lower bit of the tag.

\begin{figure}[!htb]
    \centering
    \includegraphics[height=0.29\textwidth,width=0.29\textwidth]{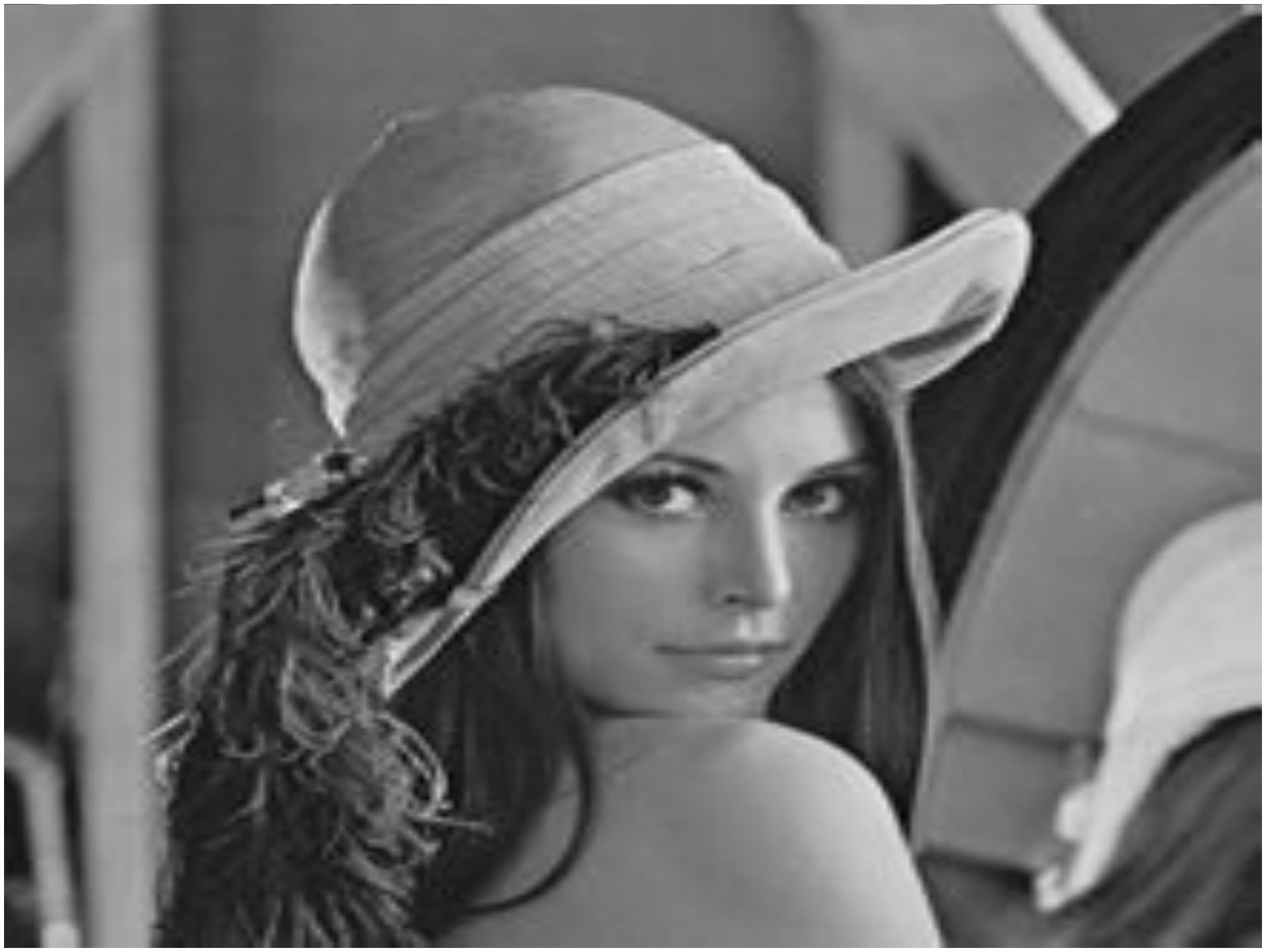}\\
    \caption{\label{fig:2-1}(a)Original lena image}
\end{figure}

The step of initial the  lena image is defined as follow;

for i = 1 : cr

  ~ for j = 1 : cc

       ~ ~ t = cover(i,j)mod 2;

          ~ ~ ~if(t == 0)

            ~ ~ ~ ~ cover(i,j)=cover(i,j)+1;

\begin{figure}[!htb]
    \centering
    \includegraphics[height=0.29\textwidth,width=0.29\textwidth]{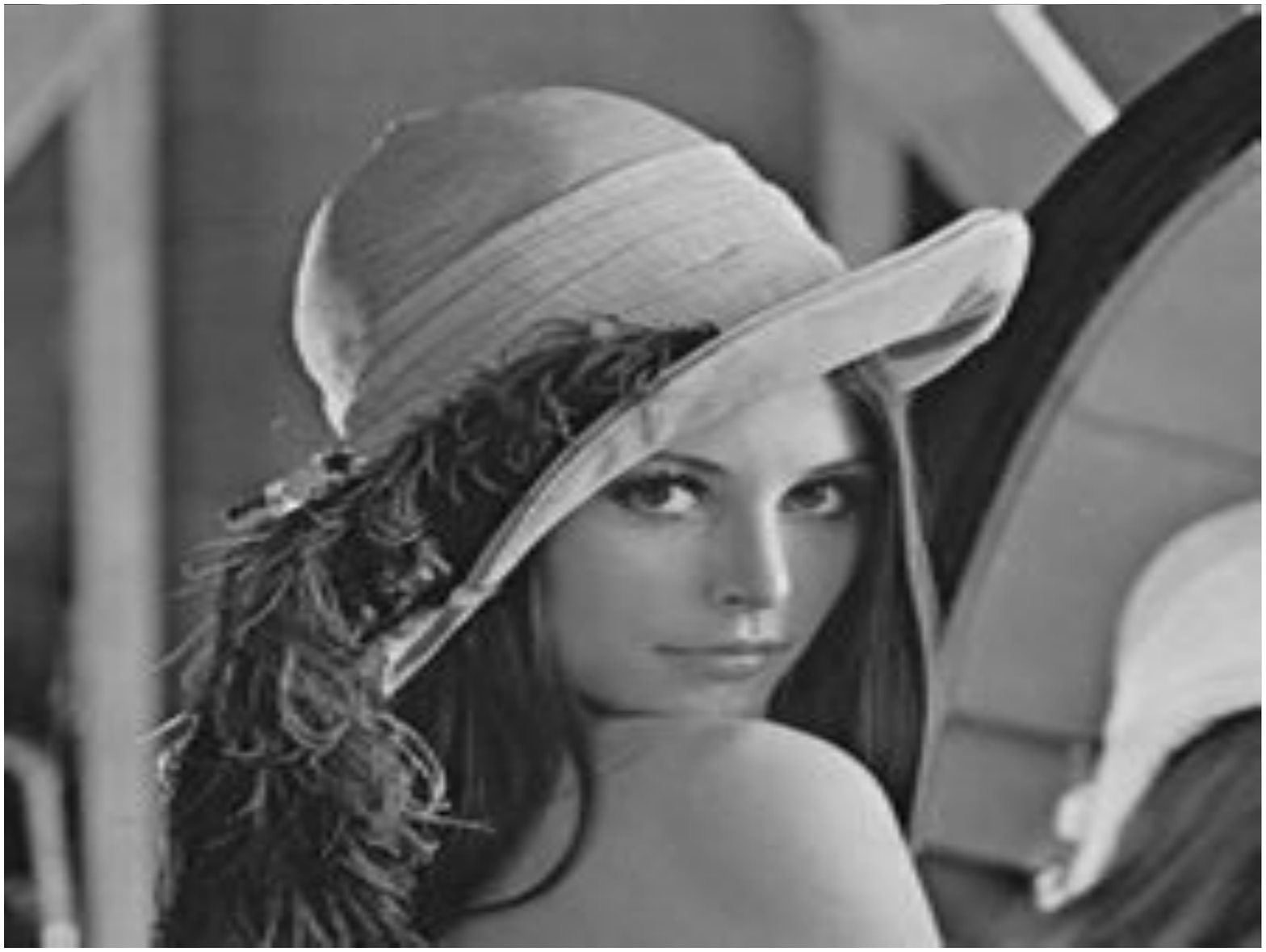}\\
    \caption{\label{fig:3-1}(b)the Initialed lena image}
\end{figure}

\begin{figure}[!htb]
    \centering
    \includegraphics[height=0.29\textwidth,width=0.29\textwidth]{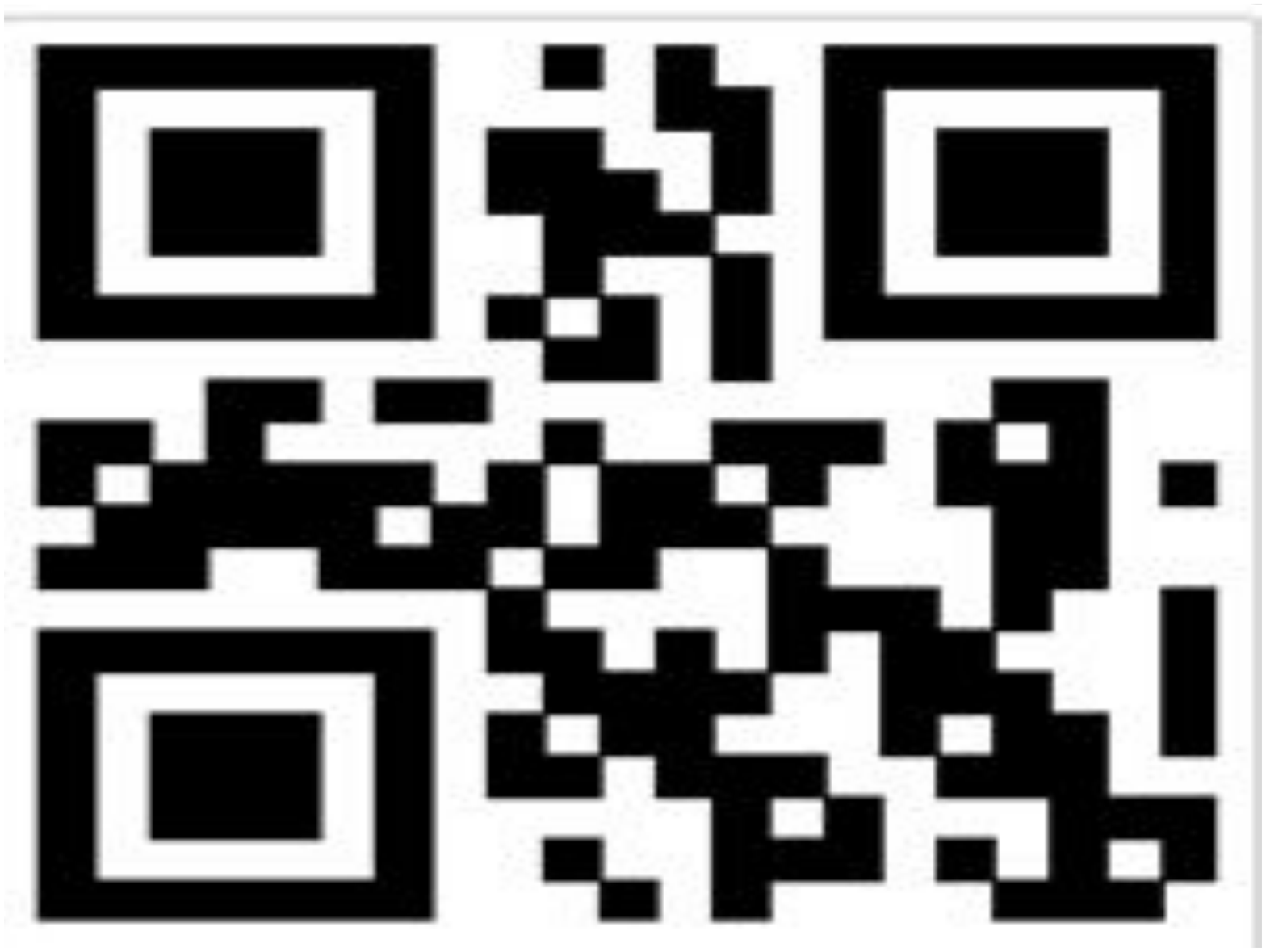}\\
    \caption{\label{fig:4-1}(c)the Original image of the tag}
\end{figure}

\begin{figure}[!htb]
    \centering
    \includegraphics[height=0.29\textwidth,width=0.29\textwidth]{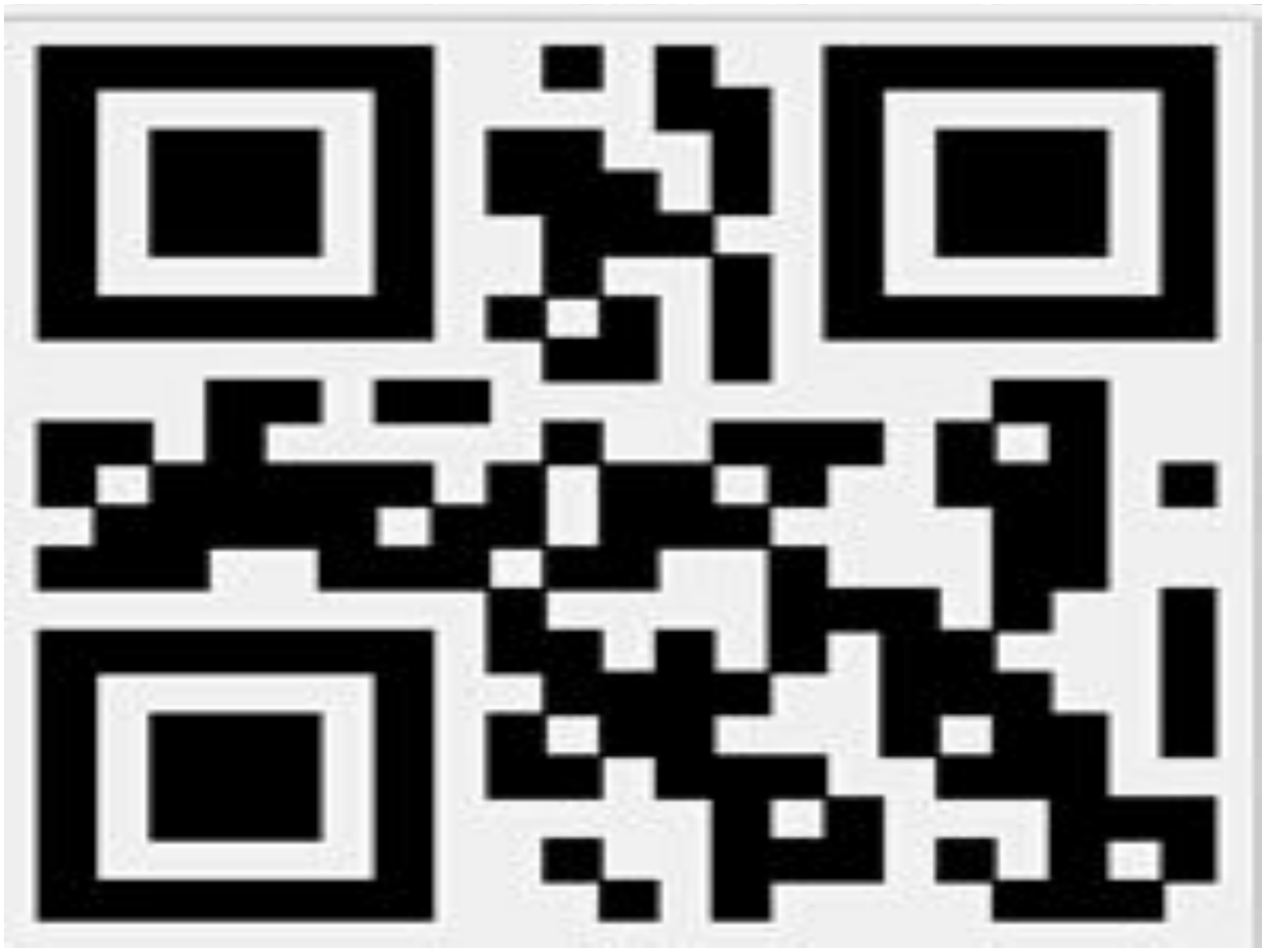}\\
    \caption{\label{fig:5-1}(d)the  Tag that lower bit become zero}
\end{figure}

\begin{figure}[!htb]
    \centering
    \includegraphics[height=0.29\textwidth,width=0.29\textwidth]{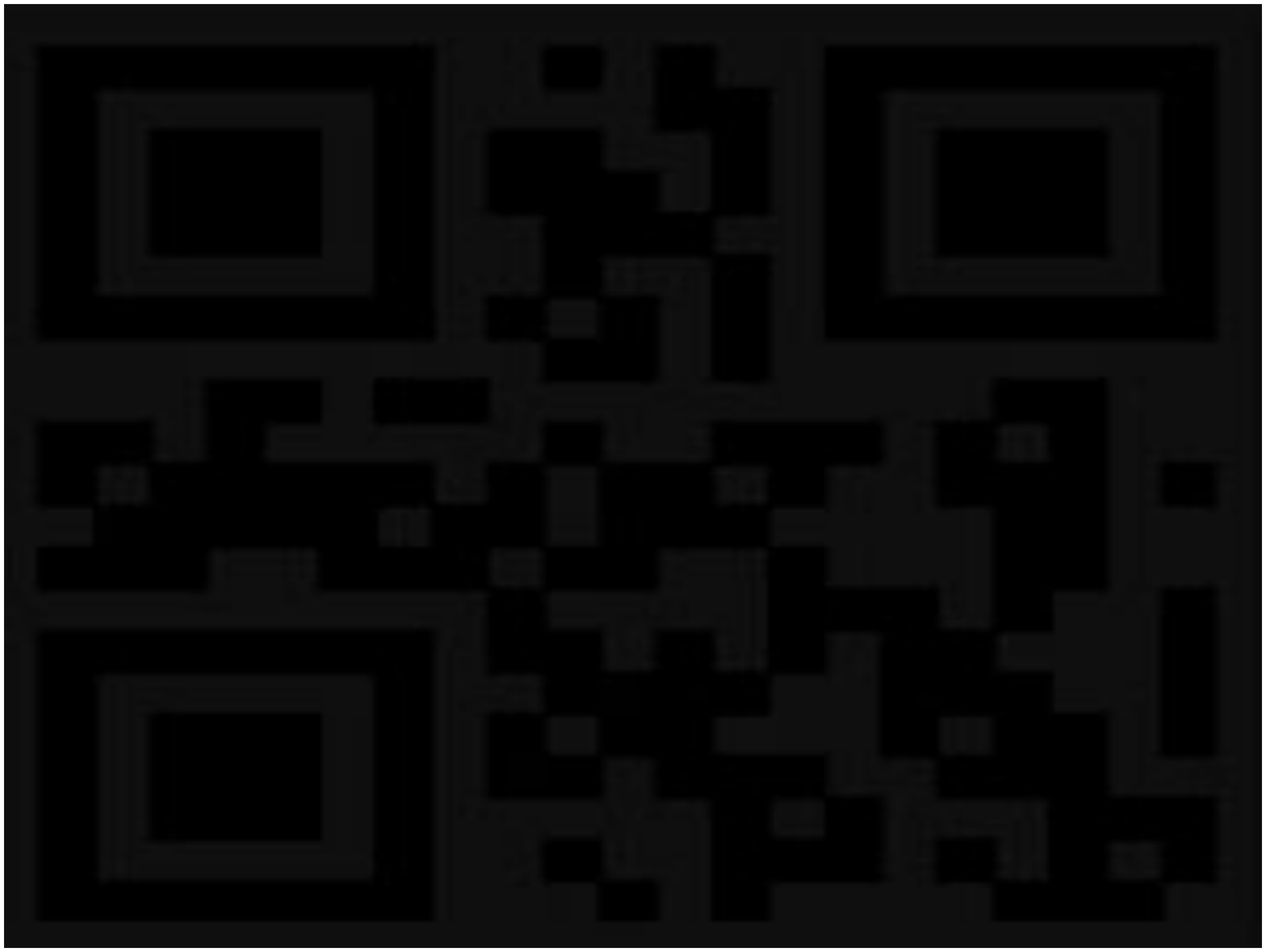}\\
    \caption{\label{fig:6-1}(e)image that the higher of tag become lower}
\end{figure}

\subsection{Decide the position to insert the tag}\label{SSEC: Decide the position to insert the tag}

We use the function rand(1,1)  to generate a random number that between 0,1 in MATLAB R2012a,every time we use a function logistic that  generate two numbers between 0,1and take some follow steps to become x, y (the index of the picture),if I find the right position x, y I make the lower Significant to zero and generate a Matrix to record  the position of the inserted tag. the detail as followed:

for i=1:mr

\hspace{4mm}for j=1:mc

     \hspace{8mm} [x,y]=logistic(k);

           \hspace{12mm}  k=x;

      \hspace{8mm} x=round(rem(x*1000,cr-1))+1;

      \hspace{8mm} y=round(rem(y*1000,cc-1))+1;

      \hspace{8mm} temp=bitand(cover(x,y),15);

      \hspace{8mm}while(temp==0)

           \hspace{12mm} x=rem(x+1,cr-1)+1;

           \hspace{12mm} temp=bitand(cover(x,y),15);

       \hspace{8mm}end

       \hspace{8mm} cover(x,y)=bitand(cover(x,y),240);

       \hspace{8mm}record(1,local)=x;

       \hspace{8mm} record(2,local)=y;

       \hspace{8mm} end

end

\begin{figure}[!htb]
    \centering
    \includegraphics[height=0.29\textwidth,width=0.29\textwidth]{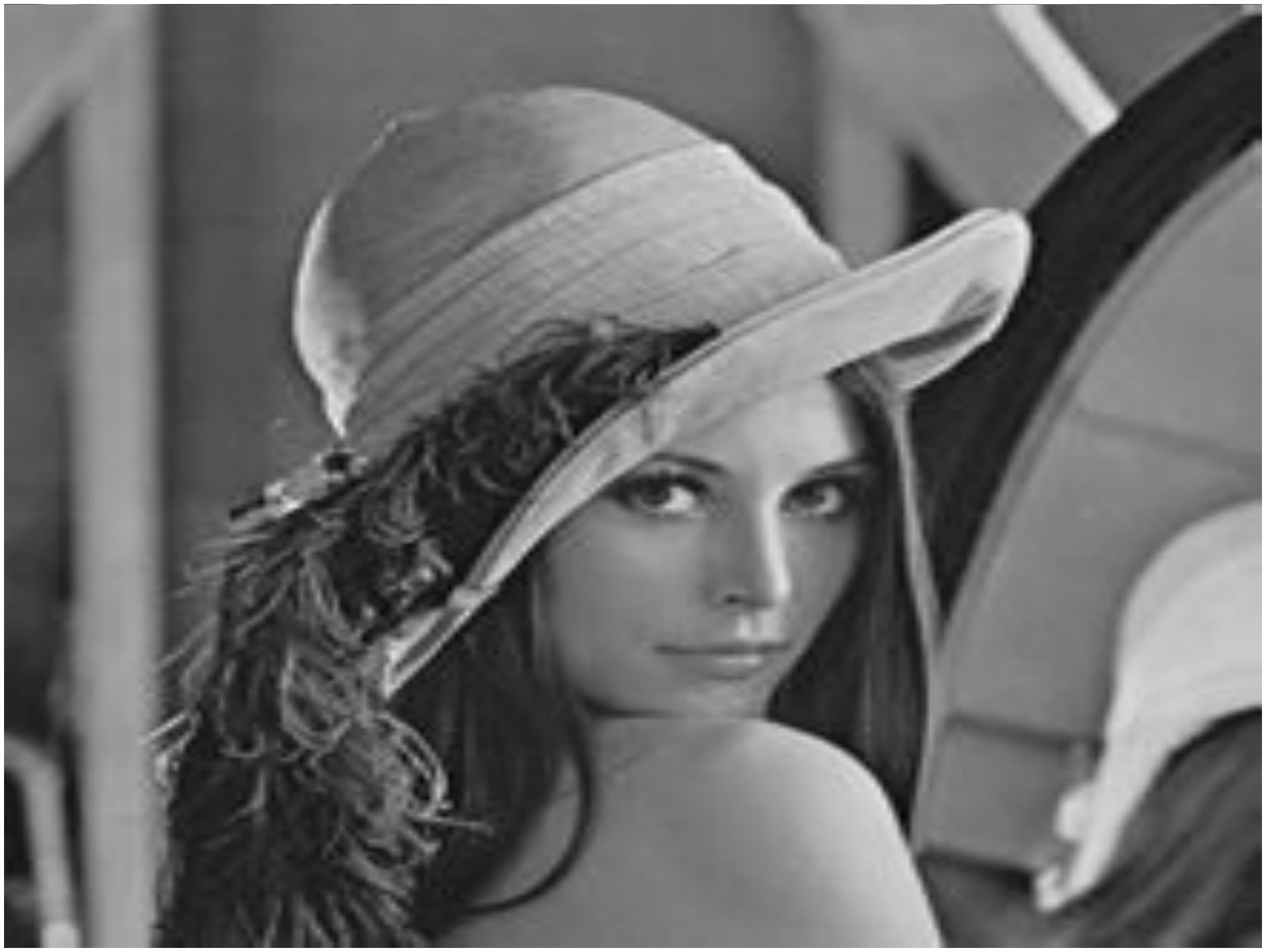}\\
    \caption{\label{fig:7-1}(a)the image that lower Significant become zero.}
\end{figure}

\subsection{Embed the tag}\label{SSEC: Embed the tag}

Once we decide where the tag should be ,we should begin to insert the information of tag, we use the way that let the higher bit of the tag replace the lower bit of the lena image, then we can make the tag invisible mostly. the simulation as followed :

\begin{figure}[!htb]
    \centering
    \includegraphics[height=0.29\textwidth,width=0.29\textwidth]{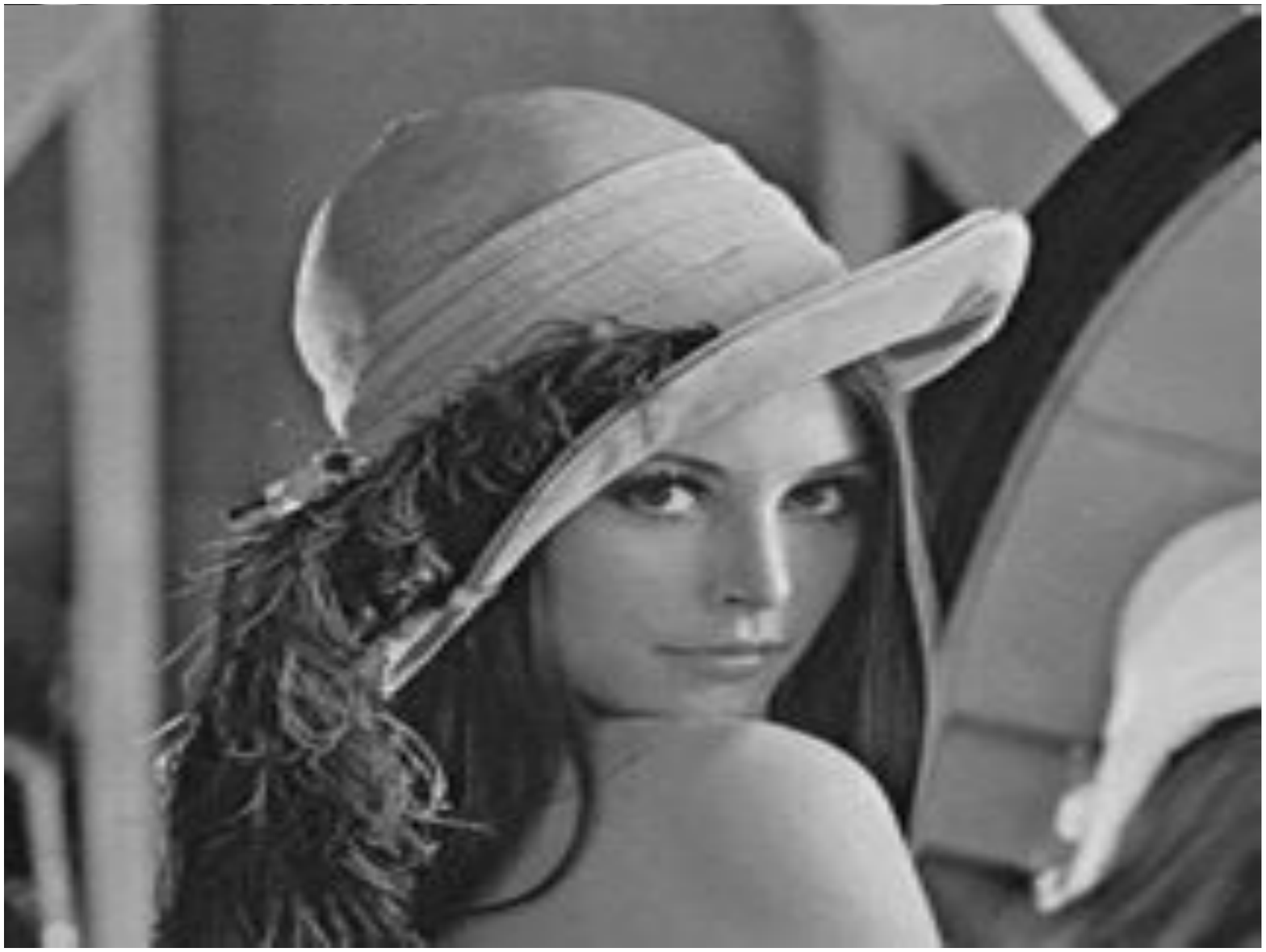}
    \caption{\label{fig:8-1}}
\end{figure}

\subsection{Extraction of the tag}\label{SSEC: Extraction of the tag}

As we said we can only use the matrix that we had made before can we ensure that we can extract the right information.
The result as followed:

\begin{figure}[!htb]
    \centering
    \includegraphics[height=0.29\textwidth,width=0.29\textwidth]{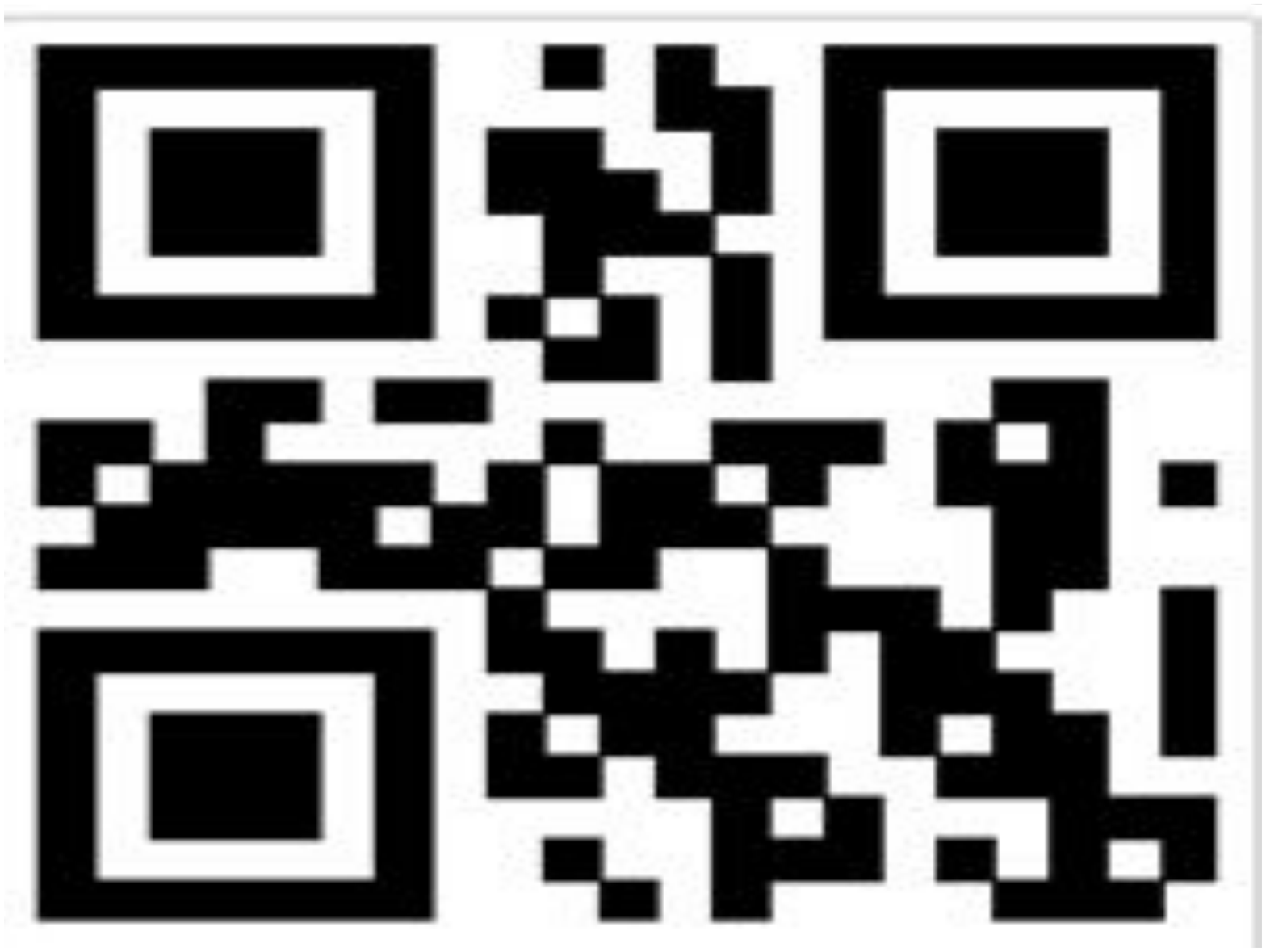}
    \caption{\label{fig:9-1}}
\end{figure}

Attack that we use a way a add some noisy in the lena that embed the tag, and use the same way to get the tag. then we get the followed result.

\begin{figure}[!htb]
    \centering
    \includegraphics[height=0.29\textwidth,width=0.29\textwidth]{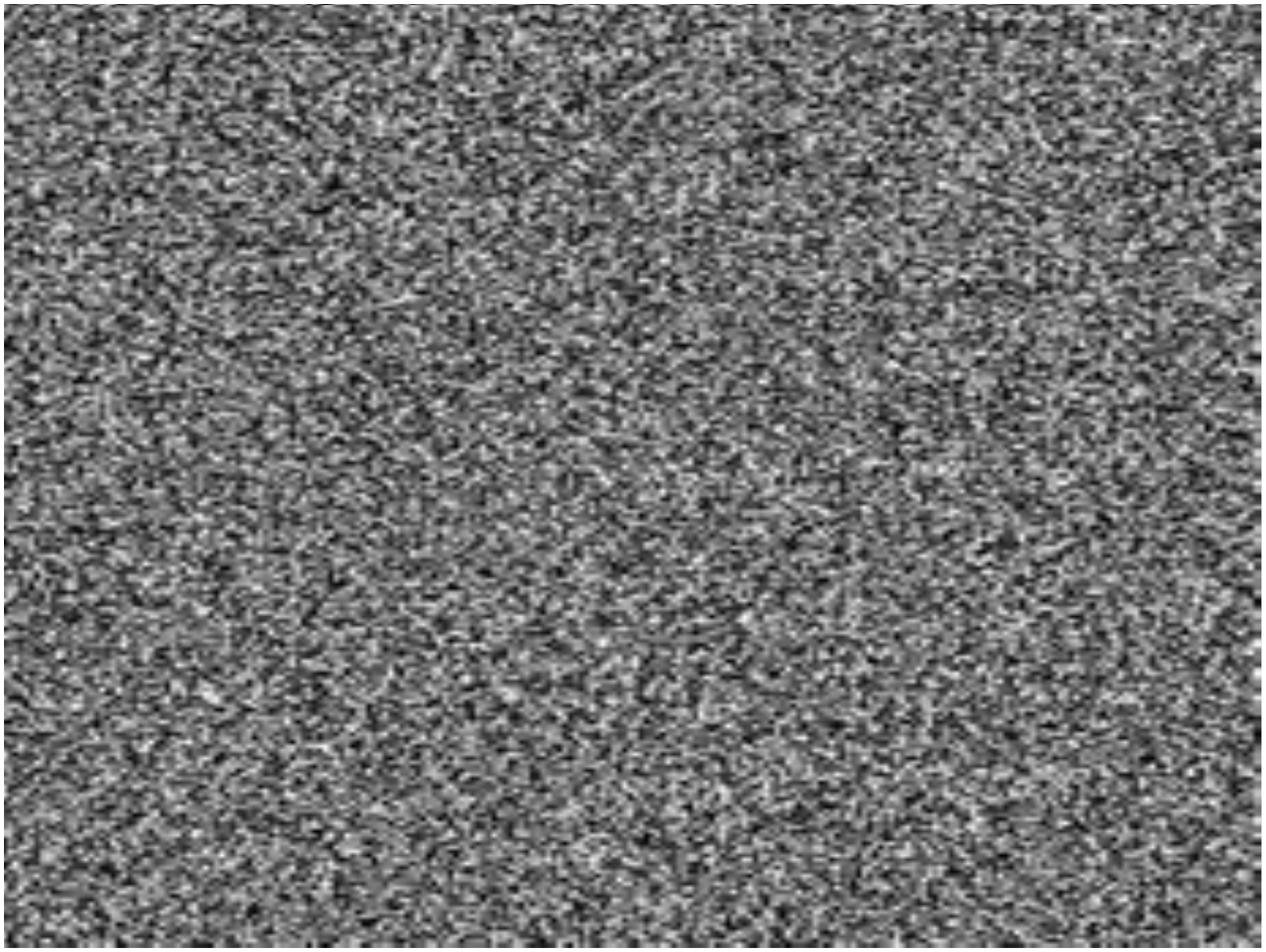}
    \caption{\label{fig:10-1}}
\end{figure}

\section{Conclusion}\label{SEC: Conclusion}

In this paper, we prove the sensitive of the picture so it can be used to  hide some information to ensure that is our product .if our tag or product be changed, we can use this way to ensure that our product and tag are connected.

\section{Acknowledgment}\label{SEC: Acknowledgment}

These research subject was supported by Sichuan University Jinjiang College, the department of Computer Science$\&$Engineering. Thanks for Prof.Bingfa Lee¡¯s suggestions and guidance.






\begin{thebibliography}{10}
 \bibitem 
 @article{eason1955certain,
  title={On certain integrals of Lipschitz-Hankel type involving products of Bessel functions},
  author={Eason, George and Noble, Benjamin and Sneddon, IN},
  journal={Philosophical Transactions of the Royal Society of London. Series A, Mathematical and Physical Sciences},
  volume={247},
  number={935},
  pages={529--551},
  year={1955},
  publisher={The Royal Society}
}
 \bibitem @book{maxwell1881treatise,
  title={A treatise on electricity and magnetism},
  author={Maxwell, James Clerk},
  volume={1},
  year={1881},
  publisher={Clarendon press}
}
 \bibitem
 @book{jacobs1963fine,
  title={Fine Particles, Thin Films, and Exchange Anisotropy:(effects of Finite Dimensions and Interfaces on the Basic Properties of Ferromagnets)},
  author={Jacobs, IS and Bean, CP},
  year={1963},
  publisher={Research Information Section, The knolls}
}
  \bibitem
  @misc{elissa2000title,
  title={Title of paper if known},
  author={Elissa, K},
  year={2000},
  publisher={unpublished}
}
   \bibitem
   @article{nicole1987title,
  title={Title of paper with only first word capitalized},
  author={Nicole, R},
  journal={J. Name Stand. Abbrev},
  pages={740--741},
  year={1987}
}
    \bibitem
    @article{yorozu1987electron,
  title={Electron spectroscopy studies on magneto-optical media and plastic substrate interface},
  author={Yorozu, T and Hirano, M and Oka, K and Tagawa, Y},
  journal={Magnetics in Japan, IEEE Translation Journal on},
  volume={2},
  number={8},
  pages={740--741},
  year={1987},
  publisher={IEEE}
}
     \bibitem
     @article{young2002technical,
  title={technical writer's handbook},
  author={Young, Matt},
  year={2002}
}
      \bibitem
      @article{voloshynovskiy2001attacks,
  title={Attacks on digital watermarks: classification, estimation based attacks, and benchmarks},
  author={Voloshynovskiy, Sviatolsav and Pereira, Shelby and Pun, Thierry and Eggers, Joachim J and Su, Jonathan K},
  journal={Communications Magazine, IEEE},
  volume={39},
  number={8},
  pages={118--126},
  year={2001},
  publisher={IEEE}
}
\bibitem
@book{cox2002digital,
  title={Digital watermarking},
  author={Cox, Ingemar J and Miller, Matthew L and Bloom, Jeffrey A and Honsinger, Chris},
  volume={53},
  year={2002},
  publisher={Springer}
}
 \end{thebibliography}
%

\citestyle{IEEEtran}
\bibliographystyle{IEEEtran}


%

\end{document}